\documentclass[preprintnumbers]{revtex4}

\input{psfig}

\usepackage{graphicx}
\def\mpl{\ifmmode \overline M_{Pl}\else $\overline M_{Pl}$\fi}

\begin{document}
\bibliographystyle{revtex}

\preprint{RU01/E-08}

\title{The Nuts and Bolts of Diffraction}

\author{Konstantin Goulianos}

\email[]{dino@physics.rockefeller.edu}
\affiliation{The Rockefeller University, New York, NY 10021\\
(Presented at ``Snowmass2001: the future of particle physics",
Snowmass, CO, USA, July 2001)
}


\begin{abstract}
Results on soft and hard diffraction
are briefly reviewed with emphasis on the interplay among
factorization properties, universality of rapidity gap formation
and unitarity.
\end{abstract}

\maketitle

\section{Hard diffraction: the question}
The signature of a diffractive event in $\bar pp$ collisions is a
leading proton or antiproton and/or a rapidity gap defined as a region of
pseudorapidity, $\eta\equiv -\ln\tan\frac{\theta}{2}$, devoid of particles.
Hard diffraction is a term used to refer to a diffractive process 
containing a hard partonic scattering
(Fig.~1). In deep inelastic scattering (DIS), diffraction is identified by
a leading proton in the final state adjacent to a rapidity gap (Fig.~2).
The rapidity gap is presumed to be due to the
exchange of a Pomeron, whose generic QCD definition is
a color-singlet combination of
quarks and/or gluons carrying the quantum numbers of the vacuum.
\begin{minipage}[t]{3.35in}
\phantom{xxx}
\vglue 0.65in
\centerline{\hspace*{0.7cm}\psfig{figure=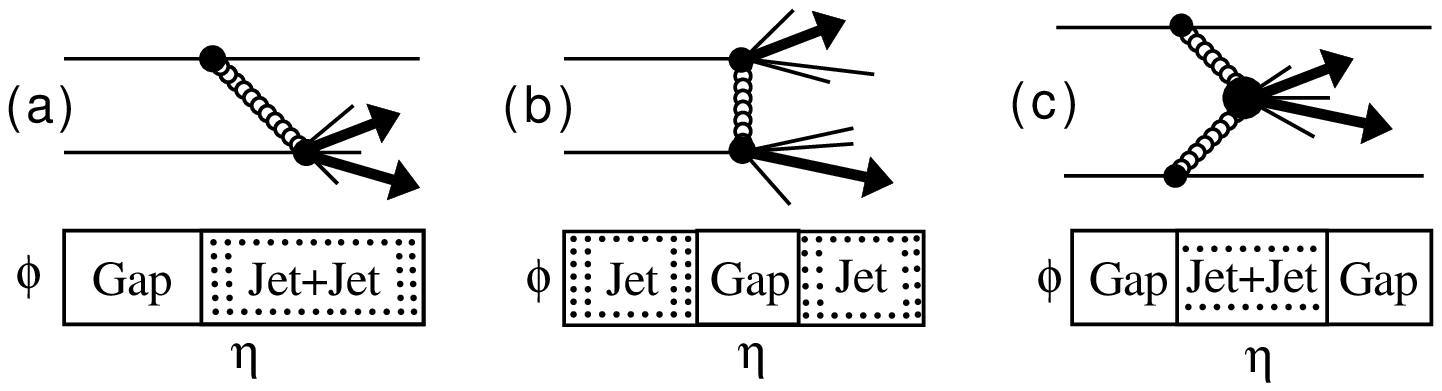,width=4in}}
\vglue -10.5cm
\vglue -3,5cm
\hspace*{1.2cm}SD\hspace*{2.2cm}DD\hspace*{2cm}DPE\\
\vglue 2.65cm
{FIG. 1: Dijet production in $\bar pp$
single (a) and double (b) diffraction, and
in double Pomeron exchange (c).}
\end{minipage}
\hfill
\begin{minipage}[t]{3in}
\phantom{xxx}
\vglue -0.6in
\centerline{\hspace*{1cm}\psfig{figure=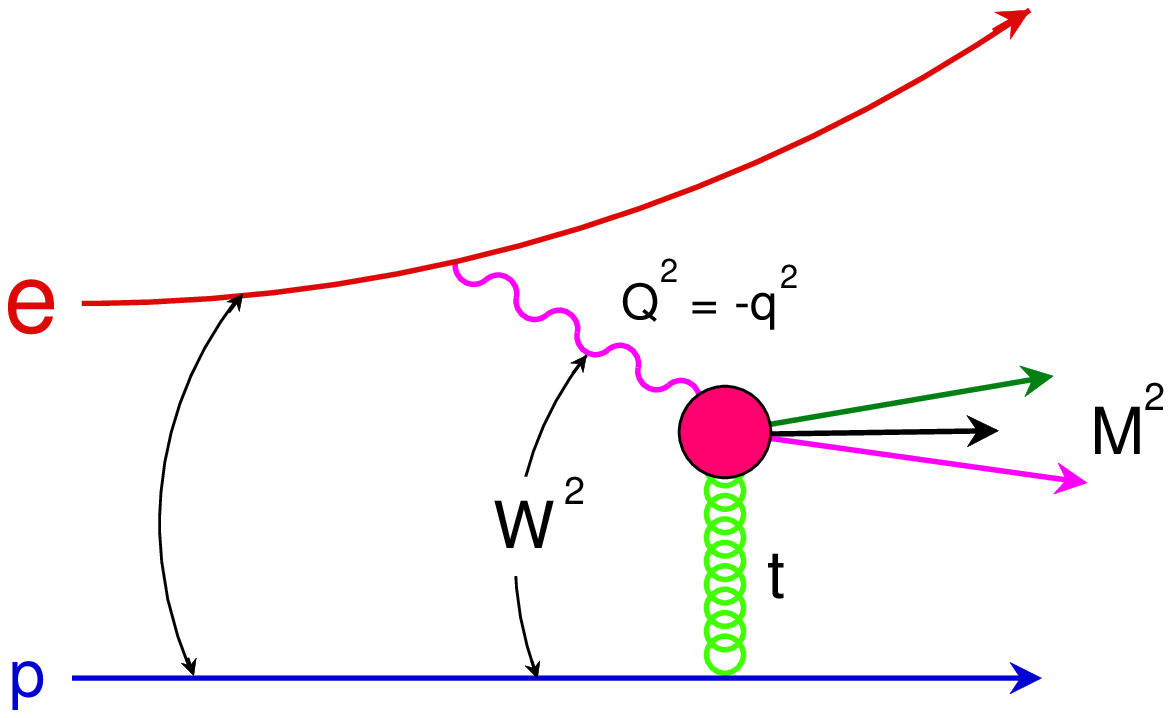,width=4in}}
\vglue -2.9in
FIG. 2: Diagram for diffractive deep inelastic scattering at HERA.
\end{minipage}

\vglue 1em
In addition to its dependence on $x$-Bjorken and $Q^2$, 
the {\em diffractive} structure function (DSF) of the leading 
nucleon could also 
depend on the nucleon's fractional momentum loss $\xi$ and its 
4-momentum transfer squared $t$.
The central question in diffraction is the validity of 
QCD factorization, i.e. whether 
hard diffraction processes can be 
described in terms of parton level cross sections convoluted with 
a universal DSF. 
\section{Data: the answer}
The question about QCD factorization in diffraction was addressed ``head on"
by a comparison~\cite{jj1800} 
between the DSF measured by CDF in dijet production 
at the Tevatron and the prediction based on parton densities extracted 
from diffractive DIS at HERA. The DSF at the 
Tevatron was found to be suppressed relative to the prediction from HERA by a 
factor of $\sim 10$. This result confirmed previous CDF results from diffractive 
$W$~\cite{W}, dijet~\cite{jj} and $b$-quark~\cite{b} production at 
$\sqrt s$=1800 GeV, and was corroborated by more recent CDF results 
on diffractive $J/\psi$~\cite{jpsi} production at 1800 GeV and 
dijet production at 630 GeV~\cite{jj630}. 

	Although factorization breaks down severely between HERA and 
the Tevatron, it nevertheless holds within the HERA data and within the 
single-diffractive data at the Tevatron at the same center of mass 
collision energy. This is demonstrated by the fact that the gluon parton 
distribution function (PDF) derived from DIS adequately describes 
diffractive dijet production at HERA~\cite{jjhera}, 
while at the Tevatron a consistent gluon PDF 
is obtained from the measured rates of diffractive $W$, dijet,
$b$-quark and $J/\psi$ production~\cite{jpsi}.  
However, factorization was found to break down at the Tevatron 
between the structure functions measured in single-diffraction and in 
double-Pomeron exchange (DPE) at $\sqrt s=$1800 GeV~\cite{DPE}.
\section{Soft diffraction: the explanation}
The breakdown of QCD factorization observed in 
hard diffraction is related to the breakdown of Regge factorization 
responsible for the suppression of soft diffraction 
cross sections at high energies~\cite{RR}. 
This may seem paradoxical, but since the rapidity gap formation is
a non-perturbative effect it should not come as a surprise. 
Thus, ``the nuts and bolts of diffraction" are contained in soft
diffraction processes.
  
Soft diffraction has been traditionally treated theoretically in the framework 
of Regge theory. For large rapidity gaps ($\Delta \eta\geq 3$), 
the cross sections for single and double (central) diffraction  
can be written as~\cite{dd}
\begin{eqnarray*}
{d^{2}\sigma_{SD}\over dtd\Delta \eta}=
  \left[{\beta^{2}(t)\over 16\pi} e^{2[\alpha(t)-1]\Delta \eta}\right]
\left[\kappa\beta^{2}(0)
{\left(\frac{s'}{\textstyle s_{\circ}}\right)}^{\alpha(0)-1}\right]
&\;\;\;&{d^{3}\sigma_{DD}\over dtd\Delta \eta d\eta_c}=
\left[{\kappa\beta^{2}(0)\over 16\pi} e^{2[\alpha(t)-1]\Delta \eta}\right]
\left[\kappa\beta^{2}(0){{\left(\frac{s'}{\textstyle s_{\circ}}\right)}}^
{\alpha(0)-1}
\right]\\
\end{eqnarray*}
\noindent where $\alpha(t)=1+\epsilon+\alpha't$
is the Pomeron trajectory, $\beta(t)$ the coupling of the
Pomeron to the (anti)proton, and $\kappa\equiv g(t)/\beta(0)$ the ratio
of the triple-Pomeron to the Pomeron-proton
couplings. 
The above two equations are remarkably similar. In each case, 
there are two terms:
\begin{itemize}
\item the first term, which is 
$\sim \{{\rm exp}[(\epsilon+\alpha't)\Delta\eta]\}^2$ and thus depends on 
the rapidity gap,
\item and the second term, which 
is $\sim$exp$\left[\epsilon\ln\left(\frac{s'}{s_0}\right)\right]$ and 
depends on the pseudorapidity 
interval $\Delta \eta'=\ln\left(\frac{s'}{s_0}\right)$ 
within which there is particle production.
\end{itemize}
In the parton model, the second term is interpreted as the sub-energy 
total cross section, while the first term is the square of the elastic
scattering amplitude between the diffractively excited state and the nucleon in 
SD, or between the two diffractive states in DD. The factor $\kappa$ 
in the second term may then 
be interpreted as being due to the color matching required for a diffractive 
rapidity gap to occur. Since the sub-energy cross section is properly
normalized, the first factor in the equations may be thought of 
as the rapidity gap probability and {\em renormalized} to 
unity. A model based on such a renormalization 
procedure~\cite{RR,KGgap} has yielded 
predictions in excellent agreement with measured SD and DD cross 
sections, as seen in Figs.~3 and 4. 

The renormalized rapidity gap
probability is by definition energy independent and thus represents 
a scaling behaviour. This procedure has the added advantage of 
preserving unitarity, which otherwise would be 
violated. Convoluting the gap probability with partonic level cross sections 
yields hard diffractive cross sections in general agreement with observations,
explaining the factorization properties discussed in the previous 
section~\cite{RR}.

\begin{minipage}[t]{3.25in}
\vspace*{-1.5cm}
\hspace*{-0.5cm}\psfig{figure=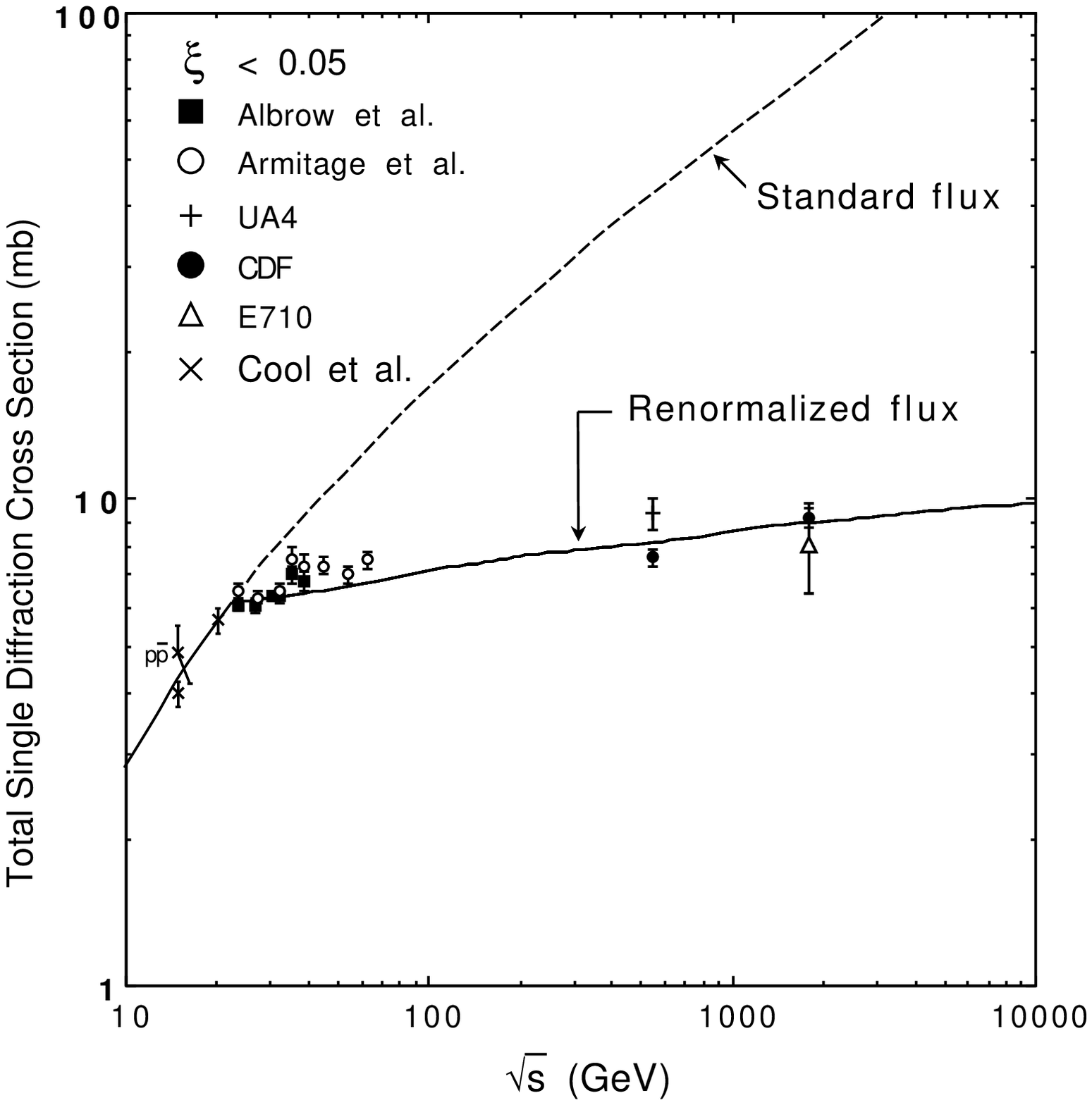,width=3.75in}
\vglue -2.9cm
FIG. 3: The $\bar pp$ total SD
cross section exhibits an $s$-dependence consistent 
with the renormalization procedure of Ref.~\cite{RR}, 
contrary to 
the $s^{2\epsilon}$ behaviour expected from Regge theory 
(figure from Ref.~\cite{RR}).
\end{minipage}
\hspace*{0.2in}
\begin{minipage}[t]{3.25in}
\vspace*{-0.4cm}
\psfig{figure=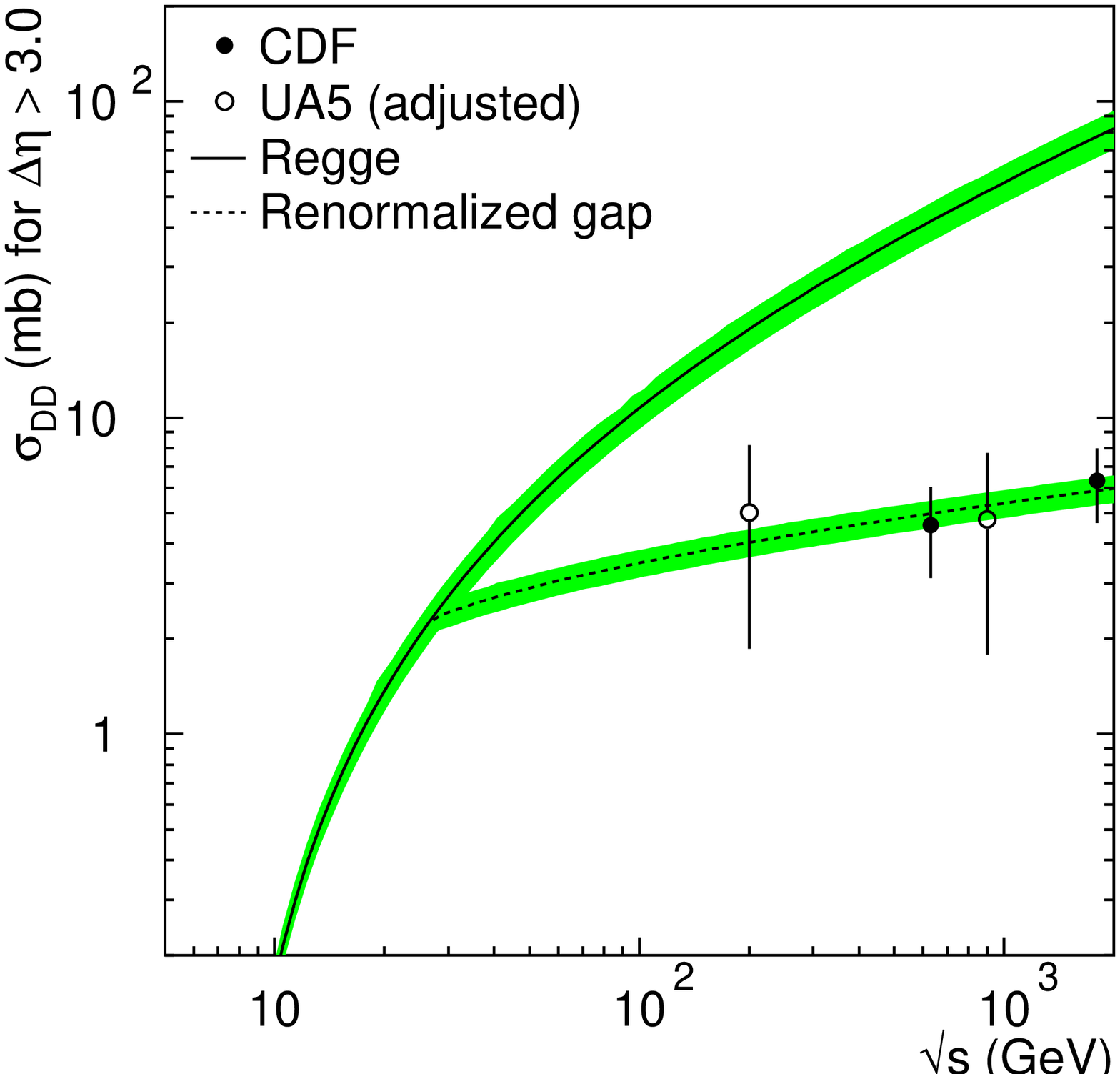,width=3.35in}
\vglue 0.05cm
FIG. 4: The $\bar pp$ total DD (central gap) cross section 
agrees with the prediction of the 
{\em renormalized rapidity gap} model~\cite{KGgap},
contrary to the $s^{2\epsilon}$ 
expectation from Regge theory (figure from Ref.~\cite{dd}). 
\end{minipage}
\newpage
\section{Multiple rapidity gaps in diffraction}
The renormalization method used to calculate the SD and DD cross sections 
can be extended to multi-gap diffractive events. Below, we outline the 
procedure for calculating the differential cross section for a 4-gap 
event (we use rapidity, $y$, and pseudorapidity, $\eta$, 
interchangeably).  
\vglue -1in
\unitlength 1in
\thicklines
\begin{center}
\begin{picture}(6,1)(1,0)
\put(1,0){\line(2,0){6}}
\multiput(2,0)(2,0){3}{\oval(0.8,0.5)[t]}
\put(1.9,-0.25){\large $y_1'$}
\put(2.9,-0.25){\large $y_2$}
\put(3.9,-0.25){\large $y_2'$}
\put(4.9,-0.25){\large $y_3$}
\put(5.9,-0.25){\large $y_3'$}
\put(2.8,0.5){\large $\Delta y_2$}
\put(3.35,-0.65){\large $\Delta y\equiv \sum_{i=1}^4\Delta y_i$}
\put(4.8,0.5){\large $\Delta y_3$}
\put(1.25,-0.9){\large $t_1$}
\put(2.9,-0.9){\large $t_2$}
\put(4.9,-0.9){\large $t_3$}
\put(6.7,-0.9){\large $t_4$}
\put(1.15,0.5){\large $\Delta y_1$}
\put(1.85,0.5){\large $\Delta y'_1$}
\put(3.85,0.5){\large $\Delta y'_2$}
\put(5.85,0.5){\large $\Delta y'_3$}
\put(6.6,0.5){\large $\Delta y_4$}
\end{picture}
\end{center}
\vglue 0.8in
\centerline{FIG. 5: Topology of a 4-gap event in pseudorapidity space.}
\vglue 1em
The calculation of the differential cross section 
is based on the parton-model scattering amplitude:
$${\rm Im\,f}(t,\Delta y)\sim e^{(\textstyle{\epsilon}+\alpha't)\Delta y}$$
For the rapidity regions $\Delta y'_i$, where there is  
particle production, the $t=0$ parton model amplitude is used and the 
{\em sub-energy cross section} is  
$\sim e^{\textstyle{\epsilon}\Delta y'}$. 
For rapidity gaps, $\Delta y$, which can be considered as resulting from 
elastic scattering between diffractively excited states, 
the square of the full
parton-model amplitude is used, 
$e^{2(\textstyle{\epsilon}+\alpha't_i)\Delta y_i}$, and the form factor 
$\beta^2(t)$ is included for a surviving nucleon.
The {\em gap probability} (product of all rapidity gap terms) 
is then normalized to unity, and 
a {\em color matching factor} $\kappa$ is included for each gap.

For the 4-gap example of Fig.~5, 
which has 10 independent variables, 
$V_i$ (shown below the figure), we have:
\begin{itemize}
\item 
$\frac{d^{10}\sigma}{\Pi_{i=1}^{10}dV_i}=
P_{gap}\times \sigma({\rm sub-energy})$
\item 
$\sigma({\rm sub-energy})=
\kappa^4\left[\beta^2(0)\cdot e^{\textstyle{\epsilon}
\Delta y'}\right]$
\hfill ($\Delta y'=\sum_{i=1}^3\Delta y'_i$)
\item 
$P_{gap}=N_{gap}\times \Pi_{i=1}^4\left[e^{(\epsilon+\alpha't_i)
\Delta y_i}\right]^2\times [\beta(t_1)\beta(t_4)]^2=N_{gap}
\cdot e^{2\textstyle{\epsilon}\Delta y}\cdot f(V_i)|_{i=1}^{10}$
\hfill ($\Delta y=\sum_{i=1}^4\Delta y_i$)\\
where $N_{gap}$ is the factor that normalizes $P_{gap}$
over all phase space to unity.
\end{itemize}
The last equation shows that the renormalization 
factor of the gap probability 
depends only on $s$ (since $\Delta y_{max}=\ln s$) 
and not on the number of diffractive gaps. Thus, the ratio of 
DPE to SD cross sections is expected to be $\approx \kappa$, 
with no additional energy dependent suppression. This can be tested 
at the Tevatron, where one may also study events with a central rapidity
gap within single-diffractive clusters. The study of events with more than 
two gaps will have to await the commissioning of the LHC.
 
\def\MPL #1 #2 #3 {Mod. Phys. Lett. {\bf#1},\ #2 (#3)}
\def\NPB #1 #2 #3 {Nucl. Phys. {\bf#1},\ #2 (#3)}
\def\PLB #1 #2 #3 {Phys. Lett. {\bf#1},\ #2 (#3)}
\def\PR #1 #2 #3 {Phys. Rep. {\bf#1},\ #2 (#3)}
\def\PRD #1 #2 #3 {Phys. Rev. {\bf#1},\ #2 (#3)}
\def\PRL #1 #2 #3 {Phys. Rev. Lett. {\bf#1},\ #2 (#3)}
\def\RMP #1 #2 #3 {Rev. Mod. Phys. {\bf#1},\ #2 (#3)}
\def\NIM #1 #2 #3 {Nuc. Inst. Meth. {\bf#1},\ #2 (#3)}
\def\ZPC #1 #2 #3 {Z. Phys. {\bf#1},\ #2 (#3)}
\def\EJPC #1 #2 #3 {E. Phys. J. {\bf#1},\ #2 (#3)}
\def\IJMP #1 #2 #3 {Int. J. Mod. Phys. {\bf#1},\ #2 (#3)}
\def\JHEP #1 #2 #3 {J. High En. Phys. {\bf#1},\ #2 (#3)}

\end{document}